\documentclass[11pt,a4paper]{article}

\usepackage[margin=1in]{geometry}
\usepackage{amsmath,amssymb,amsfonts}
\usepackage{graphicx}
\usepackage{booktabs}
\usepackage{makecell}
\usepackage{array}
\usepackage{multirow, multicol}
\usepackage{xcolor}
\usepackage{url}
\usepackage{verbatim}
\usepackage{textcomp}
\usepackage{stfloats}
\usepackage{caption}
\usepackage[caption=false,font=normalsize,labelfont=sf,textfont=sf]{subfig}

\usepackage{algorithm}
\usepackage[noend]{algpseudocode}

\usepackage[colorlinks=true, linkcolor=blue, citecolor=blue, urlcolor=blue]{hyperref}

\title{\textbf{Automated Estimation of MBIST Area and Test Time in Heterogeneous Memory IPs via Stacked Ensemble Framework}}

\author{
  Chee Jin Teoh$^{1}$, 
  Ab Al-Hadi Ab Rahman$^{1,*}$, 
  Johnny Kee Hui Wong$^{2}$, \\
  Premkumar A/L Kesavan Prabagaran$^{2}$, 
  Muhammad Nadzir Marsono$^{1}$, 
  Nuzhat Khan$^{1}$ \\[1.5ex]
  \small $^{1}$Faculty of Electrical Engineering, Universiti Teknologi Malaysia, Johor Bahru, Johor, Malaysia \\
  \small $^{2}$Intel Microelectronics (M) Sdn. Bhd., Bayan Lepas, Malaysia \\
  \small $^{*}$Corresponding author: \href{mailto:hadi@utm.my}{hadi@utm.my}
}

\date{} % Suppress date for standard preprint style

\begin{document}

\maketitle

\begin{abstract}
Embedded memories occupy a large portion of modern System-on-Chip (SoC) designs, especially in high-performance applications such as artificial intelligence and edge computing. Memory Built-In Self-Test (MBIST) is commonly used to ensure memory reliability, but it introduces additional area and test time overhead. Accurate early estimation of these overheads is important during design planning, yet conventional methods rely on full Register Transfer Level (RTL) synthesis and test pattern generation, which are slow and resource-intensive. This study proposes a supervised learning framework that predicts MBIST area and test time directly from RTL-level design parameters without synthesis. A dataset of 4,470 samples for area and 624 for test time was generated using Synopsys Design Compiler and MINT, an Intel-enhanced MBIST tool. Input features include memory count, word width, address depth, port configuration, and clock domains. For area prediction, the features are processed through polynomial expansion, log transformation, and scaling, followed by a stacked ensemble model using XGBoost, LightGBM, and a Neural Network with Gradient Boosting as the meta-learner. For test time, XGBoost and LightGBM are combined using Ridge Regression, with hyperparameters tuned through a 100-trial Optuna search. The models achieved 90.68\% accuracy for area and 96.80\% for test time within a $\pm10\%$ margin, improving over baseline methods by 8.53\% and 48.80\% respectively. The results show that this approach enables faster estimation of MBIST costs and supports more efficient design decisions in memory IP development.
\end{abstract}

\textbf{Keywords:} Memory Built-In Self-Test (MBIST), design-for-test (DFT), machine learning, stacked ensemble, feature engineering

\section{Introduction}
Memory Built-In Self-Test (MBIST) has become an indispensable component in System-on-Chip (SoC) designs, particularly as embedded memories now dominate the silicon area in modern semiconductor architectures \cite{b1}, \cite{b2}, \cite{b3}. In advanced SoCs, embedded SRAMs and register files often exceed 70\% of the total chip area \cite{b4}, \cite{b5}, necessitating efficient and scalable test strategies. Among various Design-for-Test (DFT) methodologies, MBIST accounts for the highest area overhead, reaching up to 66\% in some cases \cite{b5}, \cite{b6}, far surpassing other components like scan chains or boundary scan \cite{b7}. This underscores its central role in ensuring memory fault detection, repairability, and in-field reliability \cite{b8}, \cite{b9}. However, the integration of MBIST logic introduces significant design challenges, especially in the early stages where area overhead and test time must be balanced against stringent performance and cost requirements \cite{b5}, \cite{b10}, \cite{b11}, \cite{b12}.

\begin{figure}[!t]
\centering
\includegraphics[width=0.85\linewidth]{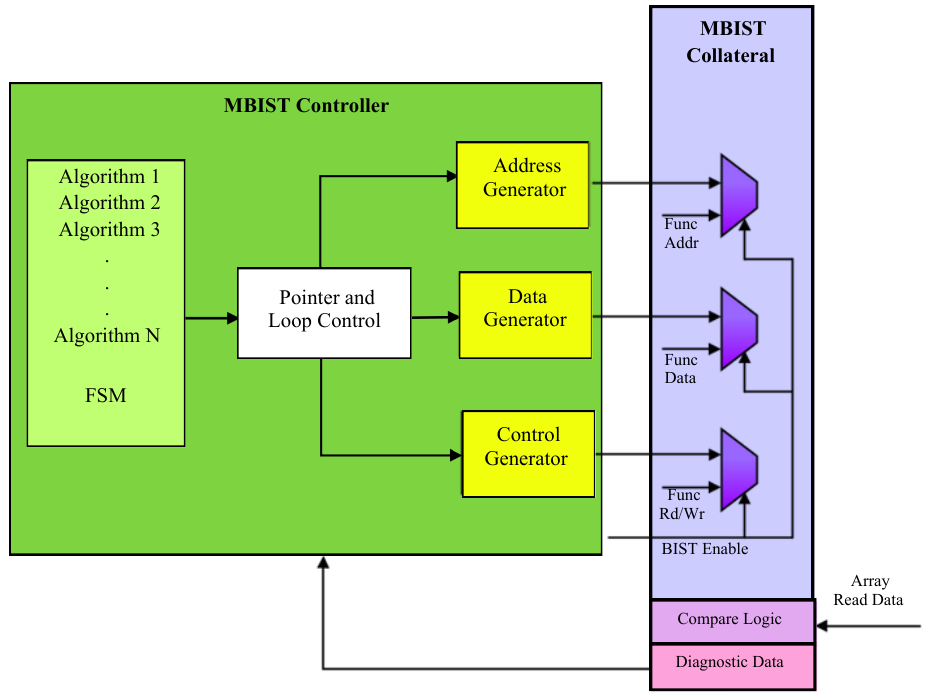}
\caption{MBIST architecture diagram that shows the MBIST Controller with its components, including the Address, Data, and Control Generators, and the MBIST Collateral for memory testing and diagnostic operations.}
\label{fig:mbistArchitecture}
\end{figure}

The architecture of MBIST, as depicted in Fig.~\ref{fig:mbistArchitecture}, exemplifies the components and process flow \cite{b9}, which forms the basis for evaluating the system's impact on overall chip performance. Despite the availability of automated solutions such as MINT, which is Intel's repackaged version of Siemens Tessent \cite{b8}, \cite{b9}, accurate estimation of MBIST area and test time remains inherently dependent on complete synthesis. Reliable metrics can only be obtained after the RTL is transformed into a gate-level netlist and test pattern generation is completed \cite{b5}, \cite{b6}, \cite{b7}, \cite{b9}, both of which are time-consuming and require significant computational resources. For complex designs containing hundreds of memory instances and diverse architectural configurations, this process may take several hours or even days. Furthermore, the iterative nature of synthesis-based workflows \cite{b10} slows design convergence and limits opportunities for early-stage exploration. The difficulty is compounded by heterogeneous RTL coding styles, inconsistent naming conventions, and varying memory topologies, all of which introduce noise and complexity into the prediction process.

Traditional analytical and statistical regression approaches often fail to capture the intricate nonlinear relationships between memory attributes, such as type, port configuration, depth, word size, and algorithm complexity, resulting impact on MBIST overhead \cite{b13}. Additionally, manual port labeling to extract key RTL signals \cite{b14} such as clocks, addresses, and data adds further bottlenecks and susceptibility to human error, limiting automation and scalability.

Recent advances in machine learning (ML) \cite{b15}, \cite{b16} have opened new opportunities for predictive modeling in Electronic Design Automation (EDA). In particular, ensemble learning methods \cite{b17}, \cite{b18} have shown promise in modeling complex, nonlinear relationships across high-dimensional datasets, while deep learning approaches \cite{b19}, \cite{b20} offer potential for pattern recognition tasks such as semantic signal classification. By combining these strengths, it becomes feasible to build a unified framework capable of early, synthesis-free estimation of MBIST area and test time across a wide range of memory configurations \cite{b21}, \cite{b22}, \cite{b23}, \cite{b24}.
A stacked ensemble learning framework \cite{b17}, \cite{b18} is proposed for predicting MBIST area and test time directly from RTL-level design parameters \cite{b25}, \cite{b26}, thus eliminating the need for synthesis or manual preprocessing. The framework integrates domain-specific feature engineering, a three-tiered regression ensemble (XGBoost, LightGBM, Neural Network), and an AI-driven RTL signal classification module \cite{b13}, \cite{b14} to provide accurate, scalable, and design-independent predictions. Validation was performed on a dataset consisting of over 5,000 memory IP variants \cite{b25}, \cite{b26}, \cite{b27}, which were automatically generated through controlled MBIST insertion and synthesis flows using MINT \cite{b8}, \cite{b9} and Synopsys Design Compiler \cite{b27}. The results highlight high prediction accuracy, accompanied by substantial reductions in runtime and development effort, making the framework highly suitable for early-stage design planning and MBIST configuration optimization.

The main contributions of this work are as follows:
\begin{enumerate}
    \item To the best of our knowledge, this is the first machine learning-based framework that applies domain-specific RTL-level feature engineering. The selected features include memory count, word width, address depth, port configuration, and the number of clock domains. These are combined with polynomial feature expansion, logarithmic transformation, and standard scaling to enable accurate estimation of MBIST area and test time directly from RTL. This eliminates the need for synthesis or test pattern generation and supports early design evaluation with lower computational cost.
    \item A stacked ensemble regression model is developed to improve prediction accuracy and generalizability across various memory configurations. The model combines XGBoost, LightGBM, and a Neural Network as base learners, with Gradient Boosting and Ridge Regression used as meta-learners for area and test time prediction respectively. This multi-model strategy captures complex data patterns and enhances robustness across datasets.
    \item An AI-based RTL signal classification module is proposed to automatically identify clock, address, data input, and data output signals based on their naming characteristics. The classifier employs character-level n-gram encoding and logistic regression, effectively removing the need for manual annotation and enabling scalable, design-independent preprocessing across diverse IP blocks.
    \item The framework is evaluated using a dataset of 4,470 samples for area and 624 samples for test time, generated with Synopsys Design Compiler and MINT. Experimental results show prediction accuracy of 90.68\% for area and 96.80\% for test time within a $\pm10\%$ margin, outperforming baseline models by 8.53\% and 48.80\% respectively, while eliminating the need for full RTL-to-GDS flows in early MBIST planning.
\end{enumerate}

\section{Related Works}
Recent research has increasingly explored the use of machine learning techniques \cite{b28}, \cite{b29}, \cite{b30}, \cite{b31} for estimating MBIST area and test time to reduce reliance on iterative synthesis flows. Initial efforts focused on regression models \cite{b29}, such as those proposed by Jamal et al. \cite{b5}, who applied linear and polynomial regression to predict MBIST overhead with reported $R^2$ values of 0.9922 for area and 0.89 for test time. Although effective in simple system-on-chip contexts, these models lacked the capacity to model non-linear dependencies and feature interactions \cite{b32}, making them less suitable for modern, feature-rich memory IP designs.

More advanced approaches have utilized ensemble methods to improve accuracy \cite{b17}, \cite{b18}, \cite{b31}. Arora et al. \cite{b7} employed models such as Random Forest, LassoCV, and Gradient Boosting Regressor \cite{b32}, \cite{b33}, \cite{b34}, \cite{b35} for MBIST area estimation. Their work demonstrated solid performance on small subsystems but showed limited generalization to larger and more diverse memory architectures \cite{b22}, \cite{b24}. The models did not incorporate stacked ensemble strategies and did not extend to test time estimation or automated preprocessing.

Other studies investigated optimization-based approaches. For example, Jamal et al. \cite{b10} integrated decision tree classifiers and K-means clustering to optimize memory groupings \cite{b21} and test strategies \cite{b32}. While beneficial for post-floorplanning design stages, these methods did not address early-stage estimation and lacked predictive capabilities that could be applied directly to RTL-level design decisions.

Despite the growing number of studies, several limitations remain. Prior work often overlooks the importance of comprehensive feature engineering \cite{b15}, suffers from limited generalization across memory types \cite{b6}, \cite{b7}, \cite{b10}, and depends heavily on manual preprocessing. To overcome these gaps, this work introduces a unified machine learning framework that combines memory-aware encodings, log and polynomial transformations, and a robust stacked ensemble model built from XGBoost, LightGBM, and Neural Networks. Furthermore, a semantic signal classification module is developed to automatically detect key RTL signals such as clock, address, and data, enabling scalable and design-independent MBIST prediction across a wide range of configurations.

\section{Problem Formulation}
The input to this problem is a collection of RTL-based memory designs \cite{b25}, \cite{b26}, denoted as $D = \{d_1, d_2, \dots, d_n\}$, where each instance $d_i$ comprises a unique configuration of embedded memories characterized by parameters such as depth, word width, port count, clock domain diversity \cite{b5}, \cite{b7}, and MBIST test algorithm selection \cite{b9}. These parameters directly influence the physical area overhead \cite{b10} introduced by MBIST insertion and the total time required for test execution.

For each design instance $d_i$, the MBIST area overhead $A_i$ and test time $T_i$ are extracted using post-synthesis and post-test pattern generation analysis via industrial EDA tools such as Synopsys Design Compiler \cite{b27} and Tessent MINT \cite{b8}, \cite{b9}. The primary objective is to estimate these two metrics early in the design cycle, without executing time-consuming synthesis or pattern generation flows.

The problem can therefore be framed as learning two functions, $f_{\text{area}}$ and $f_{\text{time}}$, that predict MBIST area and test time respectively from a design's feature representation $X_i$. Formally, this can be expressed as:
\[
A_i \approx f_{\text{area}}(X_i); \quad T_i \approx f_{\text{time}}(X_i),
\]
where $X_i$ is a domain-engineered feature vector \cite{b15} constructed from RTL-level attributes of $d_i$, including structural memory parameters and semantically classified signals such as clocks, addresses, and data.

The core challenge lies in building models that generalize well across a wide spectrum of memory configurations \cite{b36} while preserving prediction accuracy in the presence of non-linear and combinatorial interactions among the design parameters.

\subsection{Complexity of Predicting MBIST-Induced Area Overhead}
Accurately estimating MBIST area requires the model to infer complex relationships between the number, size, and configuration of memory instances \cite{b5}, \cite{b6} and the additional logic \cite{b11} introduced by MBIST insertion. Unlike generic area estimation problems, MBIST area overhead is tightly coupled with the memory test algorithm type, controller sharing strategy, and access port distribution \cite{b22}, \cite{b24}. These factors introduce non-trivial correlations and make it difficult to estimate area using simple feature sets. The learning task must capture the variability introduced by different RTL configurations \cite{b13}, \cite{b14} and produce accurate predictions before synthesis is performed.

\subsection{Modeling Temporal Behavior for Test Time Prediction}
MBIST test time estimation presents a distinct challenge, as it depends not only on memory size and count, but also on clock domains, test algorithm depth \cite{b37}, and access parallelism. The test time is influenced by how the MBIST controller sequences operations across memories \cite{b38}, \cite{b39}, especially in multi-clock environments. Estimating this metric requires reasoning over both structural and temporal attributes extracted from the design \cite{b40}. In addition, the model must generalize to unseen configurations with varying test architectures, making the prediction problem highly non-linear \cite{b41} and sensitive to design structure.

\section{Proposed Methodology}
\begin{figure}[!t]
    \centering
    \includegraphics[width=0.8\linewidth]{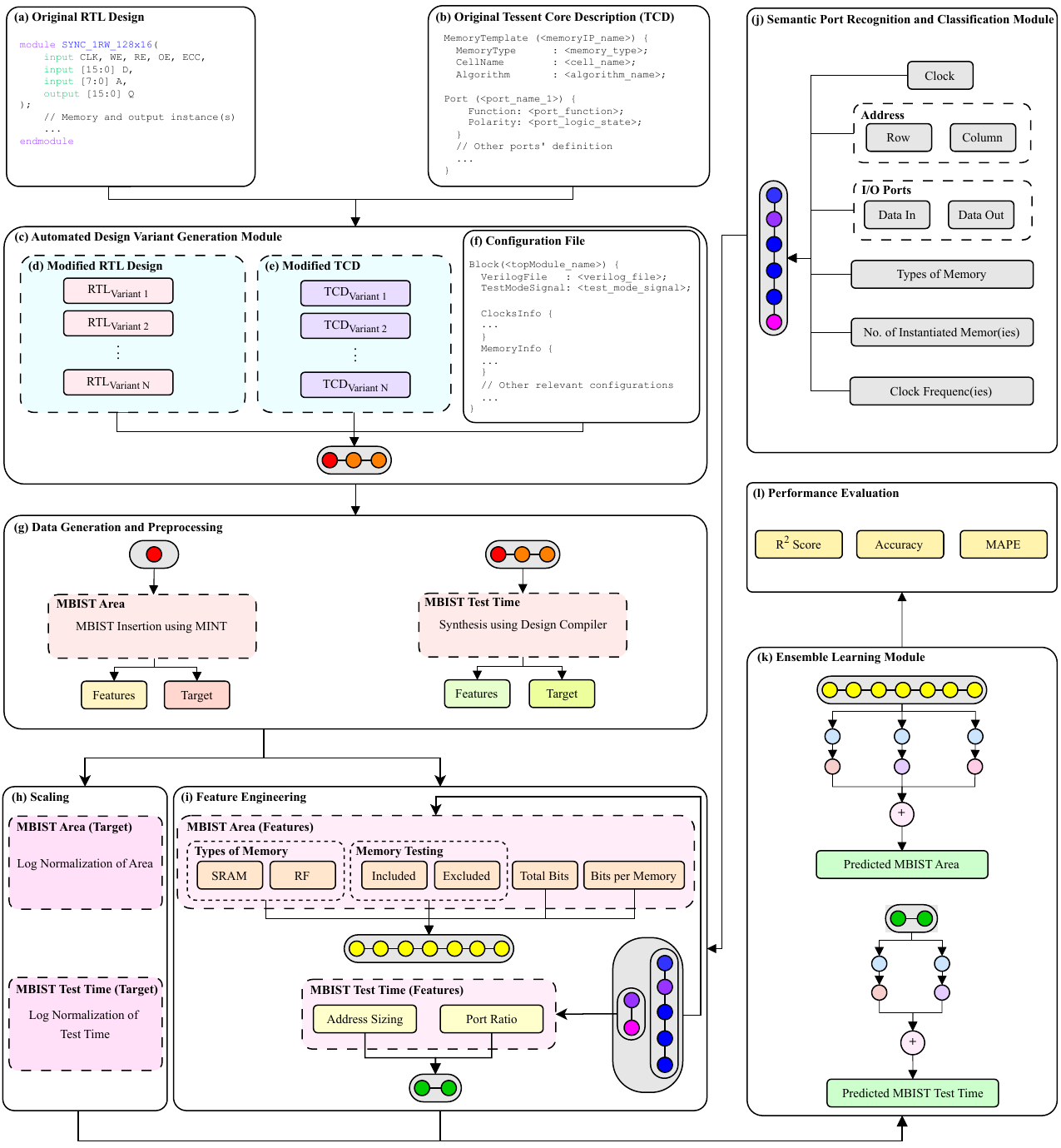}
    \caption{Overview of the proposed framework for predicting MBIST area and test time, integrating domain-specific feature engineering, stacked ensemble learning, and AI-driven signal classification for accurate and efficient prediction directly from RTL-level parameters.}
    \label{fig:overviewFramework}
\end{figure}

The proposed framework focuses on early-stage prediction of MBIST area and test time directly from RTL descriptions without invoking synthesis or test generation flows is illustrated in Fig.~\ref{fig:overviewFramework}. It begins with the automated generation of memory design variants through systematic parameter variation (see Section~\ref{sec:variant_generation}), followed by MBIST insertion and metric extraction using industrial EDA tools (see Section~\ref{sec:mbist_insertion}). Structured memory features are then extracted and transformed using polynomial expansion, logarithmic scaling, and standardization (see Section~\ref{sec:feature_engineering}). These features serve as input to a stacked ensemble learning model comprising XGBoost, LightGBM, and a Neural Network, with a gradient boosting or ridge regression meta-learner optimized via Optuna search (see Section~\ref{sec:ensemble_learning}). Additionally, a semantic signal recognition module automatically classifies RTL ports as clock, address, or data to enhance model generalizability across different coding styles (see Section~\ref{sec:semantic_port}).

\subsection{Automated Variant Generation for Dataset Construction}
\label{sec:variant_generation}
Early-stage prediction of MBIST area and test time requires a diverse and scalable dataset reflecting realistic design variations. A structured dataset was constructed by systematically modifying key architectural parameters such as memory instance count, word width, address depth, clock domain multiplicity, and the proportion of excluded memory blocks \cite{b5}, \cite{b7}. These architectural factors significantly influence controller logic complexity, clock routing challenges, and the volume of generated test patterns \cite{b10}, which in turn impact area overhead and test latency. An automation framework based on Python and GNU Makefile was employed to generate design variants by modifying RTL templates and Tessent configuration files. The resulting designs were processed through MBIST insertion using the MINT tool \cite{b8}, \cite{b9} and synthesized with Synopsys Design Compiler \cite{b27} to extract area metrics. A total of 4,470 variants were collected for MBIST area prediction and 624 for test time prediction. Each variant shares a unified core feature set and incorporates additional task-specific attributes aligned with prediction targets. Table~\ref{tab:rawfeatures} summarizes the raw features and their observed value ranges, forming a robust foundation for subsequent machine learning-based modeling.

\begin{table}[!t]
\centering
\caption{Raw Features and Corresponding Dataset Ranges Used for Predicting MBIST Area and Test Time Metrics}
\label{tab:rawfeatures}
\renewcommand{\arraystretch}{1.1}
\begin{tabular}{|p{3cm}|p{8cm}|p{3cm}|}
\hline
\textbf{Feature} & \textbf{Description} & \textbf{Range} \\
\hline
\multicolumn{3}{|c|}{\textbf{MBIST area raw features}} \\
\hline
SRAM & Number of embedded SRAM instances & 0 -- 80 \\
\hline
RF & Number of register file (RF) instances & 0 -- 80 \\
\hline
Total Memories & Total number of memories (SRAM + RF + excluded memories) & 2 -- 103 \\
\hline
Clock Domains & Number of clock domains connected to memory blocks & 1 -- 3 \\
\hline
Address Depth & Bit-width of address used to access memory & 64 -- 512 \\
\hline
Data Width & Bit-width of data bus used for read/write access & 8 -- 32 \\
\hline
\multicolumn{3}{|c|}{\textbf{MBIST test time raw features}} \\
\hline
Row Count & Bit-width of row number in the memory library & 2 -- 4096 \\
\hline
Column Count & Bit-width of column number in the memory library & 2 -- 16384 \\
\hline
Write Port & Total number of memory write ports & 1 -- 2 \\
\hline
Read Port & Total number of memory read ports & 1 -- 2 \\
\hline
\end{tabular}
\end{table}

\subsection{MBIST Insertion and Metric Extraction}
\label{sec:mbist_insertion}
Each generated design variant undergoes MBIST logic insertion using the MINT tool \cite{b8}, \cite{b9}, which performs structural validation, controller generation, pattern creation, and functional verification. During this process, MINT produces configuration files that encode the number of memory-connected clock domains and the operating frequency, both of which are essential for later test time estimation. The RTL netlist, post-insertion, is synthesized using Synopsys Design Compiler \cite{b27} to retrieve gate-level area information. MBIST area is quantified as the difference between the area of the design after insertion and the baseline area of the original RTL in \eqref{MBIST_area}:

\begin{equation}
\label{MBIST_area}
\mathit{MBIST\ Area}= \mathit{Modified\ Area}-\mathit{Original\ Area},
\end{equation}
with the relative area overhead computed as:

\begin{equation}
    \mathit{Area\ Overhead(\%)}=\frac{\mathit{MBIST\ Area}}{\mathit{Original\ Design\ Area}}\times100\%.
\end{equation}

To evaluate test time, the MINT summary report provides the total number of test cycles, which is then combined with the configured clock information. The final test time is calculated using either of the following equivalent expressions in \eqref{testTimeClockPeriod} and \eqref{testTimeClockRate}:

\begin{equation}
    \label{testTimeClockPeriod}
    \mathit{Test\ Time}= \mathit{Clock\ Period}\times\mathit{Test\ Cycles},
\end{equation}
\begin{equation}
    \label{testTimeClockRate}
    \mathit{Test\ Time}= \frac{\mathit{Test\ Cycles}}{\mathit{Frequency}}.
\end{equation}

All extracted metrics, including MBIST area, area overhead, and test time, are compiled into structured datasets alongside the corresponding architectural parameters. These datasets serve as labeled outputs for supervised machine learning models in the prediction pipeline.

\subsection{Feature Engineering and Input Transformation}
\label{sec:feature_engineering}
Feature engineering serves as a critical component in enhancing model accuracy and generalizability \cite{b15} for MBIST area and test time prediction. The transformation pipeline incorporates both raw design attributes and engineered descriptors tailored to capture complex dependencies across memory configurations. For area prediction, core inputs such as memory instance count, word width, address depth, and the number of clock domains are processed using polynomial feature expansion to capture higher-order interactions, while logarithmic scaling mitigates skewness in features with wide dynamic ranges. These transformed features are then standardized via z-score normalization to ensure consistent scaling across all samples, enabling stable learning dynamics and improved convergence behavior. The final engineered feature set, summarized in Table~\ref{tab:engineeredfeatures}, facilitates effective pattern recognition across heterogeneous memory architectures.

The test time branch expands this feature space by integrating geometry-specific parameters and access-related indicators that influence test cycle duration. Notably, attributes such as address sizing and port ratio encapsulate both structural complexity and control overhead of memory blocks. Given that test time typically correlates more linearly with structural parameters than MBIST area, the model applies fewer nonlinear transformations. Nevertheless, both the feature vectors and the log-transformed test cycle target undergo standard scaling to stabilize variance and equalize feature influence. Engineered targets used in both tasks are presented in Table~\ref{tab:engineeredtargets}.

While earlier methods often relied on minimal transformations and basic input parameters, the proposed feature construction process incorporates advanced domain-specific techniques to enrich the input space and improve model generalizability. By encoding architectural patterns, control overheads, and memory geometry into engineered features, the models are equipped to capture both linear and nonlinear dependencies inherent in MBIST behavior. This enriched representation leads to stronger learning performance, particularly when handling heterogeneous memory IPs with diverse sizes, port configurations, and test strategies. The resulting pipeline not only enhances accuracy and stability across tasks but also demonstrates scalability to unseen RTL designs that differ significantly from training distributions.

\begin{table}[!t]
\centering
\caption{Engineered Features for MBIST Area and Test Time Prediction, Focusing on Key Memory Configurations and Performance Ratios}
\label{tab:engineeredfeatures}
\begin{tabular}{|p{4cm}|p{10cm}|}
\hline
\textbf{Feature} & \textbf{Description} \\
\hline
\multicolumn{2}{|c|}{\textbf{MBIST area engineered features}} \\
\hline
Included & Indicates whether a memory is configured for MBIST insertion \\
\hline
Included Ratio & Ratio of included memories to total memory instances \\
\hline
Included per Clock Domain & Average number of included memories per clock domain \\
\hline
Memory Type & Encoded memory type (0 for SRAM, 1 for Register File) \\
\hline
Total Bits & Total memory capacity calculated as depth multiplied by width \\
\hline
Excluded Ratio & Ratio of excluded memories within the design \\
\hline
Bits per Memory & Average bit size per memory instance \\
\hline
\multicolumn{2}{|c|}{\textbf{MBIST test time engineered features}} \\
\hline
Address Sizing & Estimated number of addressable memory cells calculated from row and column \\
\hline
Port Ratio & Ratio of write to read ports with a constant added to maintain stability \\
\hline
\end{tabular}
\end{table}

\begin{table}[!t]
\centering
\caption{Engineered Targets for MBIST Metrics Prediction, Including Log-Transformed Values of Area and Test Cycles to Stabilize Variance and Improve Accuracy}
\label{tab:engineeredtargets}
\begin{tabular}{|p{4cm}|p{10cm}|}
\hline
\textbf{Target} & \textbf{Description} \\
\hline
\multicolumn{2}{|c|}{\textbf{MBIST area engineered target}} \\
\hline
Log Area & Log-transformed value of area to stabilise variance \\
\hline
\multicolumn{2}{|c|}{\textbf{MBIST test time engineered target}} \\
\hline
Log Test Cycles & Log-transformed value of test cycles used as the prediction target \\
\hline
\end{tabular}
\end{table}

\subsection{Stacked Ensemble Learning Architecture}
\label{sec:ensemble_learning}
\begin{figure}[!t]
    \centering
    \includegraphics[width=\linewidth]{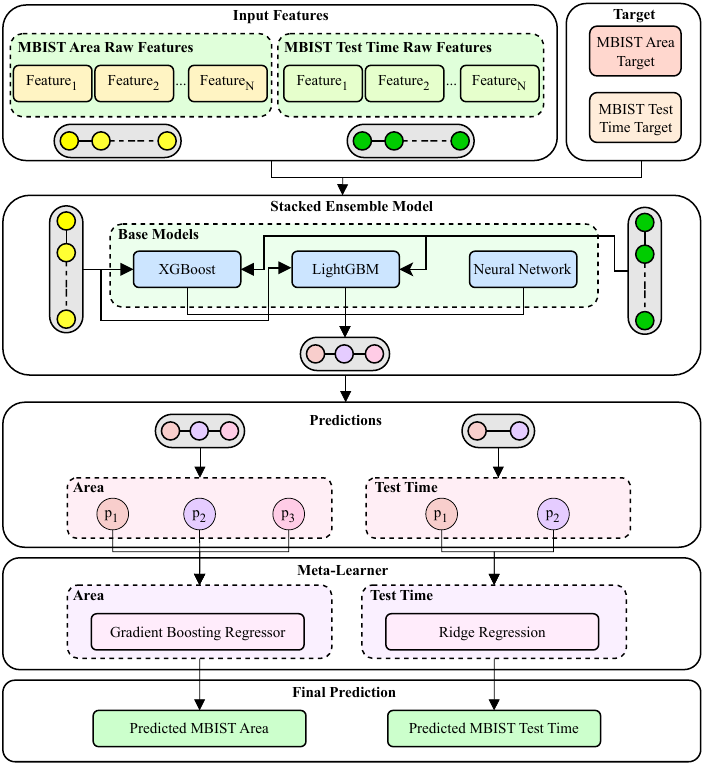}
    \caption{Proposed stacked ensemble architecture for MBIST area and test time prediction, combining base models (XGBoost, LightGBM, Neural Network) with meta-learners (Gradient Boosting and Ridge Regression) for final predictions.}
    \label{fig:ensembleArchitecture}
\end{figure}

This section presents a stacked ensemble learning architecture designed to predict MBIST area and test time from RTL-level memory configurations. The framework combines the predictive strengths of multiple base learners with a meta-learner that refines the final output \cite{b31}, \cite{b41}. Fig.~\ref{fig:ensembleArchitecture} illustrates the complete structure of the ensemble system.

In the MBIST area prediction task, the model combines three base learners, namely XGBoost, LightGBM, and a feedforward Neural Network. Each learner is trained on log-transformed and standardised features, which consist of both raw design inputs and engineered descriptors. The XGBoost and LightGBM models are well-suited for capturing hierarchical feature dependencies through gradient-boosted decision trees. The Neural Network complements them by learning complex nonlinear interactions that are difficult to express through tree-based methods. To ensure convergence and generalisation, the network employs Swish activation, dropout layers, and an early stopping mechanism based on validation performance. The predictions produced by all base models are concatenated and passed into a Gradient Boosting Regressor, which functions as the meta-learner. This regressor learns to model the residual errors of the base learners and produces the final output with improved accuracy. The full learning workflow is summarised in Algorithm~\ref{alg:area}.

For MBIST test time prediction, a more compact ensemble structure is adopted. XGBoost and LightGBM serve as the base learners, and their predictions are fused using a Ridge Regression meta-learner with L2 regularisation to improve stability. The input features are standardised, and the target variable representing test cycle count is log-transformed to reduce variance and support better convergence. The final test time is derived by scaling the predicted test cycles using the associated clock period or by applying the inverse of the operating frequency. Hyperparameters for XGBoost are optimised using a 100-trial Optuna search \cite{b30} that evaluates tree depth, learning rate, and feature sampling. LightGBM uses cross-validated defaults to maintain computational efficiency. The model training terminates once the validation MAPE satisfies a predefined convergence criterion. This workflow is outlined in Algorithm~\ref{alg:testtime}.

In contrast to prior methods that typically rely on single-model estimators or shallow feature mappings, the proposed stacked ensemble architecture demonstrates improved accuracy, scalability, and adaptability. By integrating heterogeneous learners and leveraging residual error correction through a meta-regressor \cite{b16}, \cite{b17}, the framework captures both complex nonlinear interactions and simpler linear relationships that are often overlooked by standalone models. This design also supports modular upgrades, allowing future inclusion of new model types or additional engineered features without requiring substantial changes to the overall architecture \cite{b15}. As a result, the proposed approach offers a more comprehensive and flexible solution for early-stage MBIST metric prediction.

\begin{algorithm}[!t]
\caption{Stacked Ensemble for Area Prediction}
\label{alg:area}
\begin{algorithmic}[1]
\small
\renewcommand{\algorithmicrequire}{\textbf{Input:}}
\renewcommand{\algorithmicensure}{\textbf{Output:}}
\Require 
    Dataset $D=\{X,y\}$ with engineered features, $y$ log-scaled;\\
    polynomial degree $p=2$;\\
    base learners $M_{1}$ (XGBoost), $M_{2}$ (LightGBM), $M_{3}$ (NN);\\
    meta-learner $M_{\text{meta}}$ (Gradient Boosting Regressor);\\
    convergence threshold $\varepsilon$
\Ensure 
    Trained meta-learner $M_{\text{meta}}$
\State $\tilde X \gets \text{poly\_expand}(X,p)$ \Comment{polynomial expansion}
\State $\tilde X \gets \text{standardize}(\tilde X)$ \Comment{feature scaling}
\For{$M_i \in \{M_{1},M_{2},M_{3}\}$}
    \State Train $M_i$ on $(\tilde X,y)$
\EndFor
\State $Z \gets \bigl[M_{1}(\tilde X),\,M_{2}(\tilde X),\,M_{3}(\tilde X)\bigr]$  \Comment{stacked features}
\State $\text{err}_{\text{prev}} \gets \infty$
\Repeat
    \State Train $M_{\text{meta}}$ on $(Z,y)$
    \State $\hat y \gets M_{\text{meta}}(Z)$
    \State $\text{err}_{\text{curr}} \gets \text{MAPE}(y,\hat y)$
    \If{$\lvert\text{err}_{\text{prev}} - \text{err}_{\text{curr}}\rvert < \varepsilon$}
        \State \textbf{break}  \Comment{convergence reached}
    \Else
        \State Adjust hyperparameters of $M_{\text{meta}}$
        \State $\text{err}_{\text{prev}} \gets \text{err}_{\text{curr}}$
    \EndIf
\Until{convergence}
\Return $M_{\text{meta}}$
\end{algorithmic}
\end{algorithm}

\begin{algorithm}[!t]
\caption{Stacked Ensemble for Test Time Prediction}
\label{alg:testtime}
\begin{algorithmic}[1]
\small
\renewcommand{\algorithmicrequire}{\textbf{Input:}}
\renewcommand{\algorithmicensure}{\textbf{Output:}}
\Require
    Dataset $D=\{X,y\}$ with engineered features, $y$ log-scaled;\\
    base learners $M_{1}$ (XGBoost) and $M_{2}$ (LightGBM);\\
    meta-learner $M_{\text{meta}}$ (Ridge Regression);\\
    Optuna search with $T=100$ trials;  \Comment{hyperparameter optimization}\\
    convergence threshold $\varepsilon$
\Ensure
    Optimized meta-learner $M_{\text{meta}}$
\State $X \gets \text{standardize}(X)$ \Comment{feature scaling}
\State $M_{1} \gets \text{optuna\_tune}(\text{XGBoost},X,y,T)$ 
\State Train $M_{2}$ on $(X,y)$
\State $Z \gets \bigl[M_{1}(X),\,M_{2}(X)\bigr]$   \Comment{stacked features}
\State $\text{err}_{\text{prev}} \gets \infty$
\Repeat
    \State Train $M_{\text{meta}}$ on $(Z,y)$
    \State $\hat y \gets M_{\text{meta}}(Z)$
    \State $\text{err}_{\text{curr}} \gets \text{MAPE}(y,\hat y)$
    \If{$|\text{err}_{\text{prev}}-\text{err}_{\text{curr}}| < \varepsilon$}
        \State \textbf{break}  \Comment{convergence reached}
    \Else
        \State Adjust hyperparameters of $M_{\text{meta}}$
        \State $\text{err}_{\text{prev}} \gets \text{err}_{\text{curr}}$
    \EndIf
\Until{convergence}
\Return $M_{\text{meta}}$
\end{algorithmic}
\end{algorithm}

\subsection{Semantic Port Recognition and Classification Module}
\label{sec:semantic_port}
\begin{figure}[!t]
    \centering
    \includegraphics[width=0.8\linewidth]{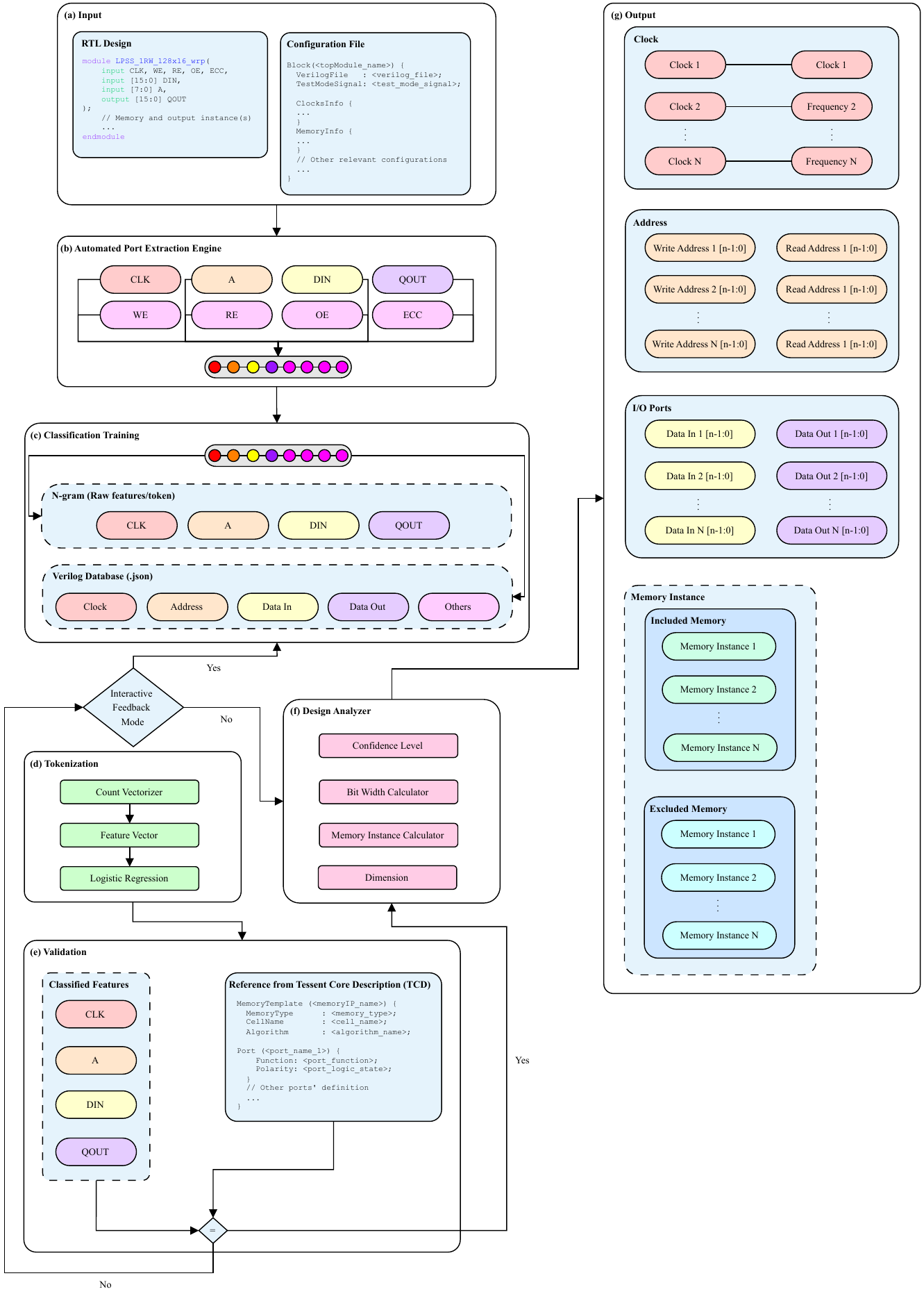}
    \caption{Proposed architecture for semantic port classification, including key components such as signal identification, classification layers, and their roles in improving RTL signal recognition.}
    \label{fig:semanticPortArchitecture}
\end{figure}

Accurate identification of signal roles such as clock, address, data input, and data output is essential for MBIST area and test time prediction. Manual labeling of these ports is often time-consuming and prone to errors, especially when naming conventions vary across RTL designs. To overcome these limitations, a semantic port recognition and classification module is introduced, as illustrated in the system architecture shown in Fig.~\ref{fig:semanticPortArchitecture}.

The process begins by parsing Verilog RTL files to extract top-level port declarations along with their corresponding identifiers. These identifiers are then tokenized using a character-level n-gram technique \cite{b13}, \cite{b14} designed to capture lexical, structural, and positional naming patterns that are often indicative of port function. The extracted token features are vectorized and passed into a logistic regression classifier trained on a labeled dataset of annotated RTL designs.

This learning-based approach allows the classifier to generalize across inconsistent naming schemes and reliably predict port categories with high accuracy. Unlike rule-based or static filtering methods, the classifier adapts to varied naming styles commonly found in both academic and industrial IP cores.

The structured CSV output, which includes signal names, bit widths, and their inferred roles, is directly integrated into the MBIST feature pipeline. This allows for consistent and automated feature extraction without requiring any manual intervention. Compared to conventional heuristics or static filtering rules, the proposed module significantly improves classification accuracy and robustness across heterogeneous IPs. Its ability to generalize across inconsistent naming schemes enables more reliable and scalable RTL analysis in complex memory IP design environments.

\section{Experimental Results and Discussion}
This section presents a comprehensive evaluation of the proposed framework for predicting MBIST area overhead and test time. The analysis is organised into five parts. Section~\ref{sec:exp_setup} describes the experimental setup, including dataset specifications and preprocessing workflow. Section~\ref{sec:hyperparam_tuning} outlines the hyperparameter optimisation strategies adopted for each learning model. Section~\ref{sec:baseline_comparison} provides a performance comparison between the proposed method and several classical baseline models. Section~\ref{sec:ablation_study} reports on ablation studies conducted to assess the contribution of ensemble learning and feature engineering. Section~\ref{sec:efficiency_tradeoff} offers a tradeoff analysis that compares the proposed method against prior works in terms of model complexity, scalability, and training efficiency.

The experiments are conducted using two structured datasets. The area dataset comprises 4470 synthesised MBIST variants, while the test time dataset contains 624 validated RTL designs. Each dataset is preprocessed through a consistent pipeline involving log transformation, polynomial feature expansion, and z-score standardisation. For area prediction, the framework employs three base learners, which are XGBoost, LightGBM, and a feedforward Neural Network. These are followed by a Gradient Boosting Regressor acting as the meta-learner. The test time prediction uses a simpler ensemble structure combining XGBoost and LightGBM, fused with a Ridge Regression meta-learner. All models are trained using the engineered feature sets described in Section~\ref{sec:feature_engineering}.

Multiple evaluation metrics are used to assess model performance. These include $R^2$ score, MAPE, and prediction accuracy within a $\pm$10\% margin. The results confirm that the stacked ensemble models outperform traditional baselines. In addition to accuracy, the framework demonstrates consistent prediction across different memory configurations. The final subsection extends the discussion by comparing the computational efficiency, training time, and architectural simplicity of this approach with previous work. This highlights the practical advantages of the proposed pipeline in terms of design-time integration and scalability for large-scale MBIST optimisation.

\subsection{Experimental Setup}
\label{sec:exp_setup}
The experimental evaluation is based on two large datasets generated through an automated flow that processes RTL designs into measurable MBIST metrics. The first dataset includes 4470 samples containing MBIST area information obtained after synthesis, while the second consists of 624 validated samples for test time prediction, extracted following test pattern generation. MBIST insertion is performed using the MINT framework \cite{b8}, \cite{b9}, and synthesis is conducted using Synopsys Design Compiler \cite{b27} on a Linux-based server environment. This setup ensures consistency and scalability in design processing across all variants.

Model development and training are conducted on a Windows 11 workstation equipped with an Intel Core i7-1370P processor and 16 GB of RAM. Python 3.11 is used as the programming platform, alongside libraries such as scikit-learn, XGBoost, LightGBM, TensorFlow, and Optuna. The data is divided into training and testing subsets using an 80 to 20 ratio. Validation is internally applied during Neural Network training and hyperparameter optimisation to maintain generalisation performance. Model accuracy is evaluated using multiple metrics including the coefficient of determination ($R^2$ score), MAPE, and prediction accuracy within a $\pm$10\% error margin from the actual values as defined below \cite{b41}.

\begin{align}
    R^2 \text{ Score} &= 1 - \frac{\sum_{i=1}^{n}(y_i - \hat{y}_i)^2}{\sum_{i=1}^{n}(y_i - \bar{y})^2} \label{eq:r2} \\
    \text{MAPE} &= \frac{1}{n} \sum_{i=1}^{n} \left|\frac{y_i - \hat{y}_i}{y_i}\right| \times 100\% \label{eq:mape} \\
    \text{Accuracy} &= \frac{1}{n} \sum_{i=1}^{n} \mathbf{1} \left( \frac{|y_i - \hat{y}_i|}{y_i} \leq 0.10 \right) \times 100\% \label{eq:accuracy}
\end{align}

The $y_i$ represents the actual values, $\hat{y}_i$ represents the predicted values, $\bar{y}$ is the mean of the actual values, and $n$ is the total number of data points. The $R^2$ score quantifies the proportion of the variance in the actual values explained by the model’s predictions, with higher values indicating better model performance. MAPE measures the average percentage error between the predicted and actual values by calculating the absolute difference between $y_i$ and $\hat{y}_i$, normalized by $y_i$, and then multiplying by 100. Lower MAPE values indicate better accuracy. The accuracy metric calculates the percentage of predictions that fall within a 10\% error margin of the true values, with the indicator function $\mathbf{1}(\cdot)$ returning 1 when the condition is met and 0 otherwise. Together, these metrics provide a comprehensive evaluation of the model's prediction accuracy, with $R^2$ reflecting goodness of fit, MAPE providing error magnitude, and accuracy assessing how often predictions are within an acceptable range \cite{b16}.

\subsection{Hyperparameter Tuning}
\label{sec:hyperparam_tuning}
Hyperparameter tuning plays an essential role in enhancing the generalisation performance of machine learning models. It involves optimising key parameters that influence model behaviour during training, including learning rate, tree depth, and regularisation strength for boosting models, as well as network architecture and dropout configuration for Neural Networks. This study employs the Optuna framework to automate the tuning process for both MBIST area and test time prediction tasks. All tuning is performed on training partitions after appropriate feature transformation, using fixed random seeds to ensure reproducibility.

For the MBIST area prediction task, a stacked ensemble model composed of XGBoost, LightGBM, and a feedforward Neural Network is developed. Input features are processed using polynomial expansion and standardisation \cite{b15}, \cite{b19}. The XGBoost and LightGBM models are tuned with respect to their maximum depth, learning rate, number of estimators, subsample ratio, and regularisation terms. Typical values explored include tree depths from 4 to 12 and estimator counts ranging from 500 to 2000. The Neural Network model is constructed with multiple dense layers, Swish activations, batch normalisation, and dropout layers. Its learning rate and training epochs are optimised using early stopping to avoid overfitting. After tuning, outputs from all base learners are combined and fed into a Gradient Boosting Regressor meta-learner, which is further optimised using a quantile loss objective \cite{b34}, \cite{b35}. Table~\ref{tab:area_hyperparams} summarises the final hyperparameters for each component.

\begin{table}[!t]
\centering
\caption{Tuned Hyperparameters for MBIST Area Prediction across Single Base Learners}
\label{tab:area_hyperparams}
\begin{tabular}{l|c|c|c}
\toprule
\textbf{Hyperparameter} & \textbf{XGBoost} & \textbf{LightGBM} & \textbf{Neural Network} \\
\midrule
Max Depth & 7 & 7 & -- \\
Learning Rate & 0.0135 & 0.0130 & 0.0015 \\
N Estimators & 850 & 850 & -- \\
Subsample & 0.92 & 0.92 & -- \\
Colsample Bytree & 0.87 & 0.87 & -- \\
Reg Alpha & 0.15 & 0.15 & -- \\
Reg Lambda & 0.35 & 0.35 & -- \\
Dropout Rate & -- & -- & 0.3 / 0.2 \\
Batch Size & -- & -- & 128 \\
\bottomrule
\end{tabular}
\end{table}

For MBIST test time prediction, the ensemble model consists of XGBoost and LightGBM as base learners, with Ridge Regression serving as the meta-learner. The features are standardised, and domain-specific engineered attributes such as port ratios and memory geometry are included. XGBoost is optimised using 100 trials of Optuna’s Tree-structured Parzen Estimator \cite{b30}, targeting mean absolute error on the log-transformed test cycle count. LightGBM uses hyperparameters refined from cross-validation experiments. The Ridge Regression model uses L2 regularisation with fixed alpha to stabilise blending of predictions from the base models. The tuning process results in improved convergence, reduced validation error, and enhanced prediction accuracy. The selected hyperparameters are listed in Table~\ref{tab:testtime_hyperparams}.

\begin{table}[!t]
\centering
\caption{Tuned Hyperparameters for MBIST Test Time Prediction across Single Base Learners}
\label{tab:testtime_hyperparams}
\begin{tabular}{l|c|c}
\toprule
\textbf{Hyperparameter} & \textbf{XGBoost} & \textbf{LightGBM} \\
\midrule
Max Depth & 4 & 10 \\
Learning Rate & 0.0278 & 0.015 \\
N Estimators & 719 & 1500 \\
Subsample & 0.65 & 0.80 \\
Colsample Bytree & 0.85 & 0.85 \\
Gamma & $1.083 \times 10^{-4}$ & -- \\
Reg Alpha & $7.397 \times 10^{-4}$ & 0.05 \\
Reg Lambda & $1.400 \times 10^{-4}$ & 0.5 \\
\bottomrule
\end{tabular}
\end{table}

\subsection{Baseline Methods and Performance Comparison}
\label{sec:baseline_comparison}
This section presents a comprehensive comparison between the proposed stacked ensemble framework and various baseline models used for MBIST area and test time prediction. Classical machine learning methods such as Linear Regression (LR), Polynomial Regression (PolyReg), LassoCV, AdaBoostRegressor, Gradient Boosting Regressor (GBR), K-Nearest Neighbors (KNN), and Neural Network (NN) were implemented with and without feature engineering. These baselines serve as a foundation to evaluate the gains introduced by advanced ensemble learning and domain-specific transformations.

In the case of MBIST area prediction, the simple regression models exhibited weak performance, especially in capturing nonlinear relationships across parameters such as SRAM instance count, address and data widths, and port configurations. Polynomial Regression marginally improved over Linear Regression, but still suffered from high error margins. The more expressive models like GBR and NN offered better accuracy but encountered overfitting issues due to the increasing model complexity. As shown in Table~\ref{tab:area_metrics}, even the strongest baseline (NN with feature engineering) reached only 82.15\% accuracy. In contrast, the proposed stacked ensemble, combining XGBoost, LightGBM, and NN as base learners with a GBR meta-learner, achieved a test accuracy of 90.68\% and an $R^2$ score of 0.9999. This 8.53\% improvement demonstrates the ensemble's ability to capture residual patterns more effectively while preserving generalisation.

For MBIST test time prediction, baseline methods faced even greater challenges due to the intricate relationship between test cycles, clock frequency, and MBIST algorithm configuration. Without feature engineering, models like LR and KNN performed poorly with high MAPE values exceeding 1000\%. Feature engineering improved the predictive power of GBR and NN, but their accuracy remained limited. The proposed method, using XGBoost and LightGBM fused through Ridge Regression, achieved a substantial leap to 96.80\% test accuracy and reduced MAPE to 2.10\%, outperforming the best baseline (GBR with feature engineering) by 48.80\% as seen in Table~\ref{tab:testtime_metrics}. The Ridge meta-model effectively mitigated variance and overfitting in the final prediction.

\begin{figure}[!t]
  \centering
  \subfloat[]{
    \includegraphics[width=0.47\linewidth]{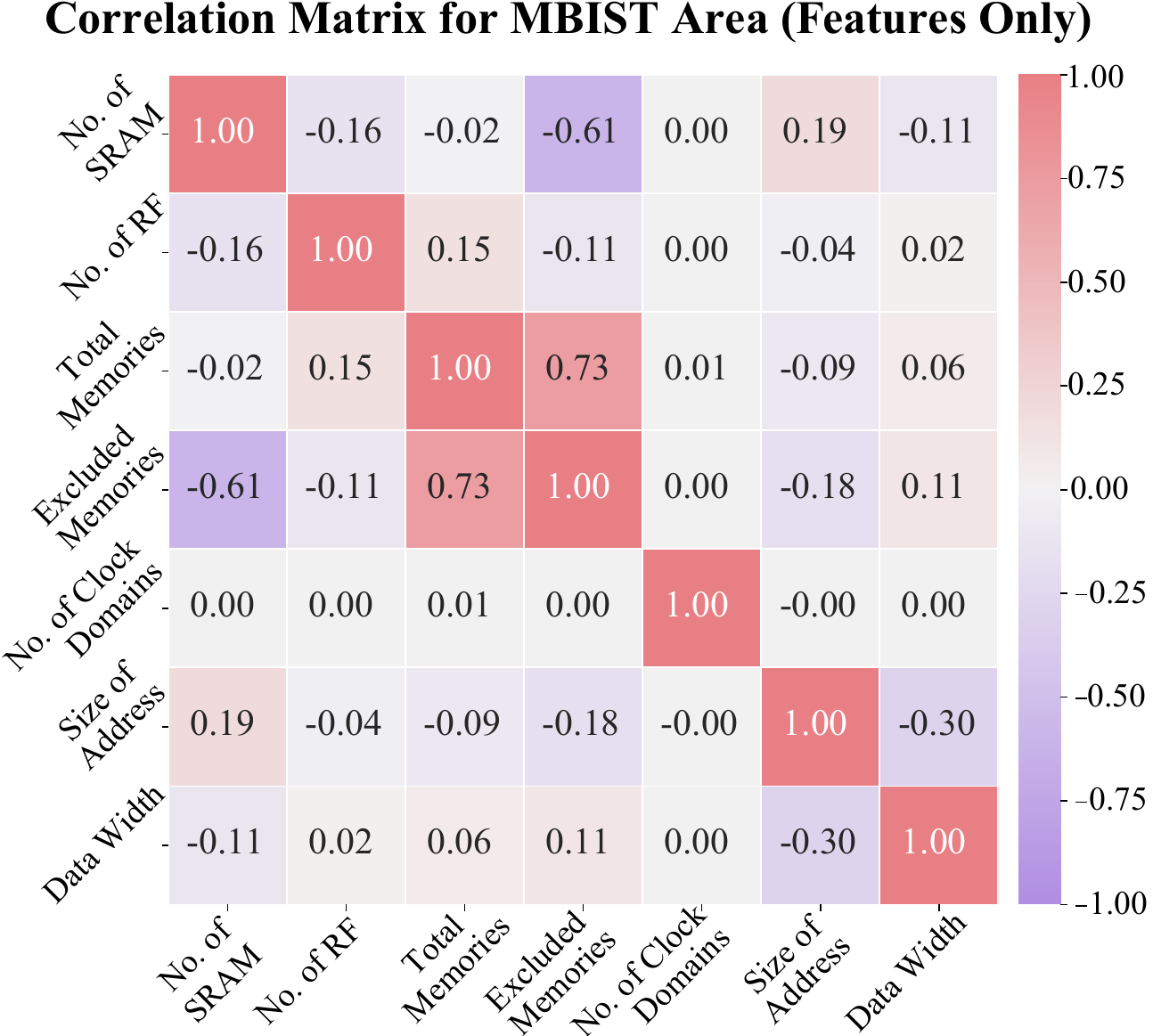}
  }
  \hfill
  \subfloat[]{
    \includegraphics[width=0.47\linewidth]{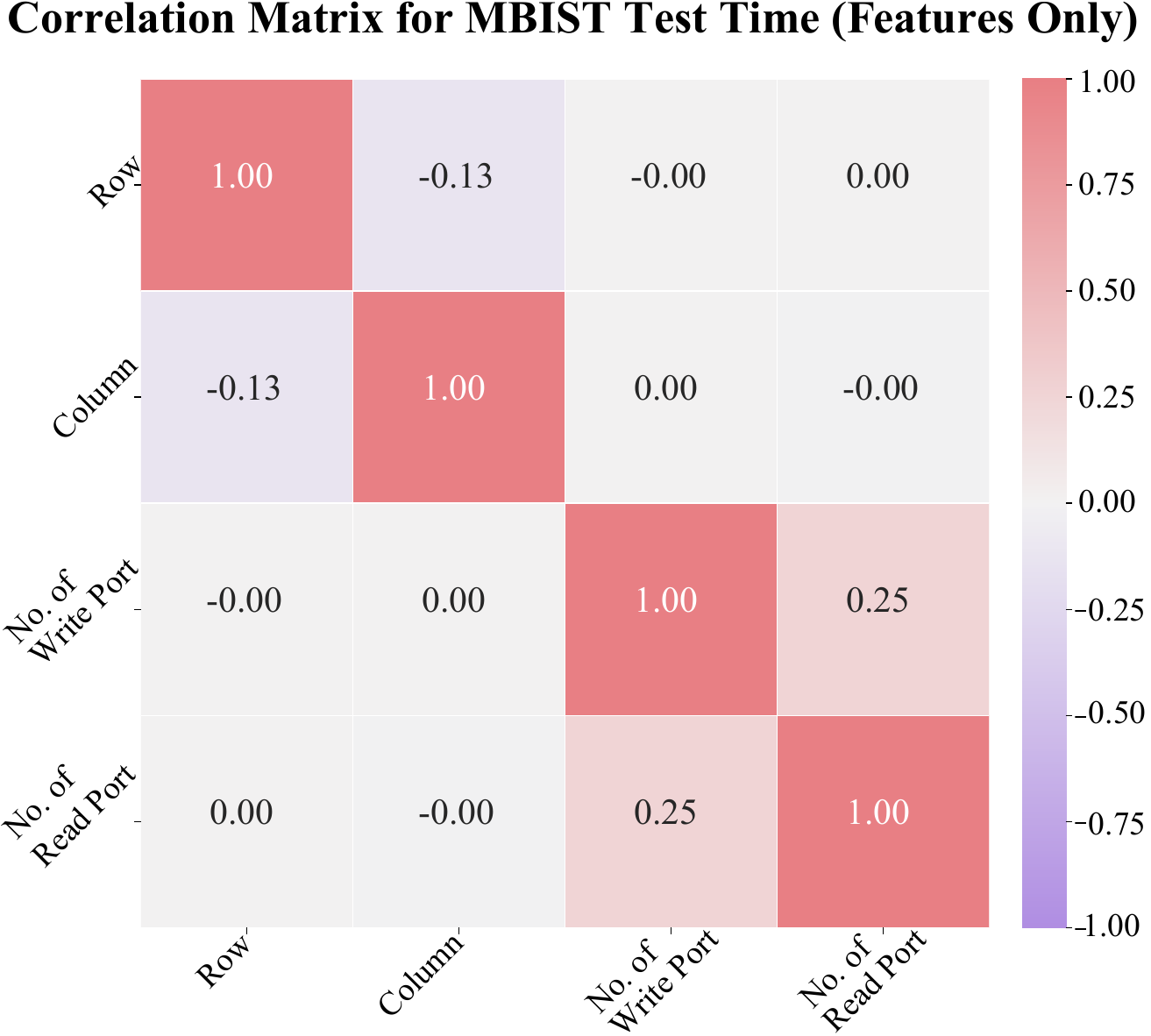}
  }
  \caption{Correlation matrix showing the relationships between key design parameters and MBIST (a) area and (b) test time, where diagonal values should be as high as possible to indicate true positive relations, and off-diagonal values should be low to indicate minimal correlations.}
  \label{fig:correlation_matrix}
\end{figure}

The correlation matrices for MBIST area and test time features, shown in Figure~\ref{fig:correlation_matrix}, highlight key relationships between the design parameters. For MBIST area, a strong positive correlation ($\sim 0.73$) between the total number of memories and excluded memories indicates that higher memory counts lead to more memories being excluded from MBIST. The number of SRAM instances shows a moderate negative correlation ($\sim -0.61$) with excluded memories, suggesting that more SRAM reduces the number of excluded memories, as SRAM blocks are prioritized for MBIST. Other features like the number of clock domains and data width show little correlation with total memories but still influence MBIST area through bus widths and controller complexity. For MBIST test time, the correlation matrix shows that row and column counts are strongly related, meaning larger memory arrays require longer test times. The moderate correlation ($\sim 0.25$) between write and read port counts reflects their role in test concurrency, which affects test scheduling. These findings underscore the importance of feature engineering to capture these relationships for accurate predictions.

These results validate that the stacked ensemble architecture, coupled with advanced feature engineering, is critical to achieving robust and scalable MBIST estimation. The use of tree-based models ensures efficiency and strong generalisation, while Neural Networks capture high-order interactions. Logarithmic and polynomial feature transformations resolve non-linear design dependencies that traditional models fail to learn. The framework significantly advances prediction accuracy across complex and heterogeneous RTL designs, confirming its superiority over existing standalone methods.

\begin{table*}[!t]
\centering
\caption{Comparison of MBIST Area Prediction Metrics With and Without Feature Engineering.}
\label{tab:area_metrics}
\resizebox{\textwidth}{!}{%
\begin{tabular}{ll ccc ccc}
\toprule
\multirow{2}{*}{\textbf{\makecell{Feature\\Engineering}}} 
& \multirow{2}{*}{\textbf{Model}}
& \multicolumn{3}{c}{\textbf{Training Set}} 
& \multicolumn{3}{c}{\textbf{Test Set}} \\
\cmidrule(lr){3-5} \cmidrule(lr){6-8}
& & \textbf{R\textsuperscript{2} Score} & \textbf{Accuracy (\%)} & \textbf{MAPE (\%)} 
& \textbf{R\textsuperscript{2} Score} & \textbf{Accuracy (\%)} & \textbf{MAPE (\%)} \\
\midrule
\multirow{7}{*}{No} 
& Linear Regression \cite{b5} & 0.4074 & 6.10 & 2979.99 & 0.3756 & 4.26 & 2871.76 \\
& Polynomial Regression \cite{b5} & 0.8254 & 19.41 & 1973.07 & 0.7927 & 18.07 & 2066.40 \\
& LassoCV \cite{b7} & 0.4074 & 6.12 & 2982.90 & 0.3757 & 4.26 & 2874.53 \\
& AdaBoostRegressor \cite{b7} & 0.9886 & 46.97 & 137.83 & 0.9882 & 45.79 & 137.44 \\
& Gradient Boosting Regressor \cite{b7,b10} & 0.9999 & 71.32 & 22.65 & 0.9998 & 70.59 & 21.07 \\
& Neural Network \cite{b7} & 0.9943 & 68.06 & 20.69 & 0.9947 & 68.24 & 16.70 \\
& Proposed Work & 0.9971 & 75.64 & 73.33 & 0.9946 & 69.13 & 121.00 \\
\midrule
\multirow{7}{*}{Yes} 
& Linear Regression \cite{b5} & 0.9854 & 51.83 & 36.62 & 0.9875 & 50.51 & 34.65 \\
& Polynomial Regression \cite{b5} & 0.9289 & 46.74 & 17.31 & 0.9327 & 45.90 & 15.78 \\
& LassoCV \cite{b7} & 0.8855 & 23.40 & 2319.90 & 0.8740 & 21.21 & 2190.96 \\
& AdaBoostRegressor \cite{b7} & 0.9887 & 47.39 & 117.50 & 0.9884 & 46.35 & 108.76 \\
& Gradient Boosting Regressor \cite{b7,b10} & 0.9994 & 80.25 & 20.07 & 0.9994 & 79.91 & 22.18 \\
& Neural Network \cite{b7} & 0.9990 & 83.43 & 10.37 & 0.9992 & 82.15 & 8.95 \\
& \textbf{Proposed Work} & \textbf{1.0000} & \textbf{91.60} & \textbf{5.54} & \textbf{0.9999} & \textbf{90.68} & \textbf{5.14} \\
\bottomrule
\end{tabular}%
}
\end{table*}

\begin{table*}[!t]
\centering
\caption{Comparison of MBIST Test Time Prediction Metrics With and Without Feature Engineering.}
\label{tab:testtime_metrics}
\resizebox{\textwidth}{!}{%
\begin{tabular}{ll ccc ccc}
\toprule
\multirow{2}{*}{\textbf{\makecell{Feature\\Engineering}}} 
& \multirow{2}{*}{\textbf{Model}}
& \multicolumn{3}{c}{\textbf{Training Set}} 
& \multicolumn{3}{c}{\textbf{Test Set}} \\
\cmidrule(lr){3-5} \cmidrule(lr){6-8}
& & \textbf{R\textsuperscript{2} Score} & \textbf{Accuracy (\%)} & \textbf{MAPE (\%)} 
& \textbf{R\textsuperscript{2} Score} & \textbf{Accuracy (\%)} & \textbf{MAPE (\%)} \\
\midrule
\multirow{4}{*}{No} 
& Linear Regression \cite{b5} & 0.2761 & 4.68 & 2795.41 & 0.1302 & 6.78 & 1909.91 \\
& Gradient Boosting Regressor \cite{b5} & 0.8327 & 39.48 & 16.49 & 0.7818 & 33.60 & 19.40 \\
& K-Nearest Neighbors \cite{b5,b10} & 0.4412 & 4.80 & 88.15 & 0.4412 & 4.80 & 88.15 \\
& Proposed Work & 0.9936 & 35.87 & 532.88 & 0.9360 & 22.40 & 788.14 \\
\midrule
\multirow{4}{*}{Yes} 
& Linear Regression \cite{b5} & 0.9053 & 7.89 & 2062.17 & 0.8989 & 7.02 & 1138.62 \\
& Gradient Boosting Regressor \cite{b5} & 0.9520 & 50.50 & 12.21 & 0.9095 & 48.00 & 14.18 \\
& K-Nearest Neighbors \cite{b5,b10} & 0.9072 & 19.84 & 23.76 & 0.8315 & 19.20 & 35.26 \\
& \textbf{Proposed Work} & \textbf{0.9997} & \textbf{100.00} & \textbf{1.06} & \textbf{0.9978} & \textbf{96.80} & \textbf{2.10} \\
\bottomrule
\end{tabular}%
}
\end{table*}

\subsection{Ablation Study on Stacked Ensemble and Feature Engineering}
\label{sec:ablation_study}
Ablation experiments were conducted to evaluate the relative impact of each component within the proposed stacked ensemble framework, focusing on MBIST area and test time prediction. The objective was to examine how individual elements such as feature engineering and specific base learners contribute to the overall prediction performance. The ensemble integrates XGBoost, LightGBM, and a Neural Network, while incorporating domain-driven feature transformations including polynomial expansion and logarithmic scaling. The evaluation considered the effect of selectively removing these components on model accuracy.

\begin{figure}[!t]
  \centering
  \subfloat[]{
    \includegraphics[width=0.47\linewidth]{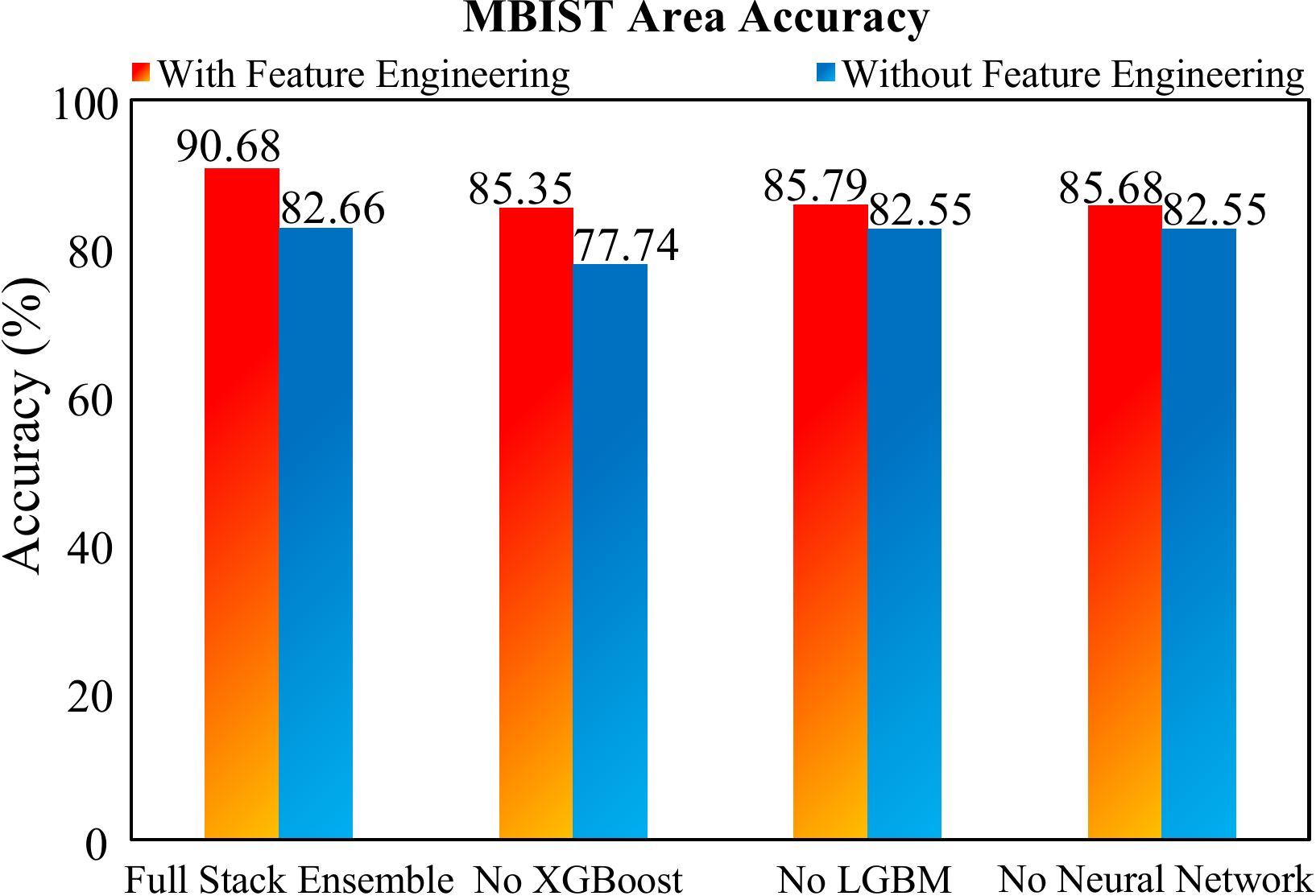}
  }
  \hfill
  \subfloat[]{
    \includegraphics[width=0.47\linewidth]{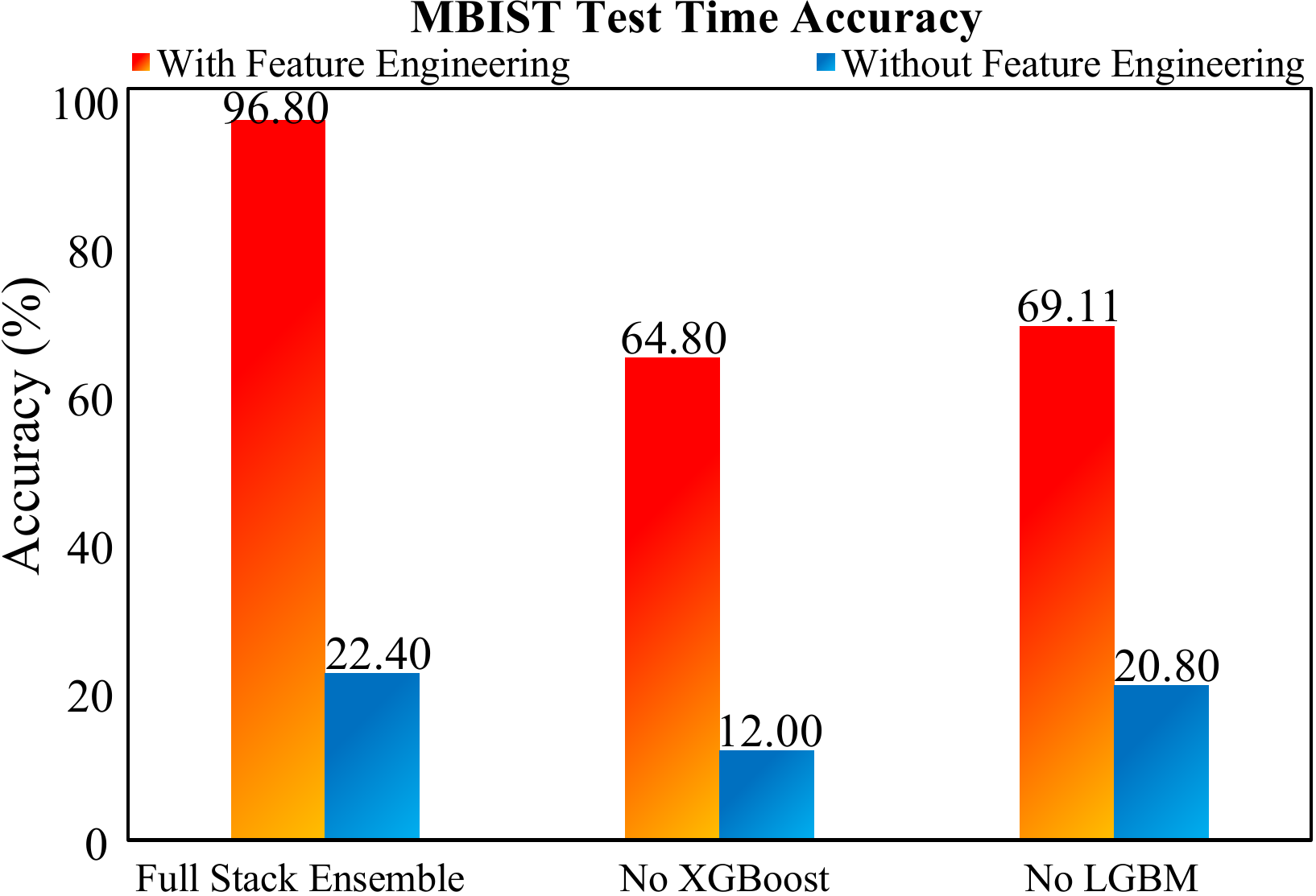}
  }
  \caption{Comparison of MBIST area and test time accuracy in the ablation study, illustrating the effects of both feature engineering and base learners on model performance for (a) area and (b) test time.}
  \label{fig:ablationmetric}
\end{figure}

For MBIST area prediction, the complete ensemble achieved a maximum test accuracy of 90.68\%, as presented in Fig.~\ref{fig:ablationmetric}(a). Removing any of the base learners resulted in a measurable decline in performance, indicating that each model captured distinct data patterns essential for robust learning. Tree-based models such as XGBoost and LightGBM provided effective handling of nonlinear boundaries and interactions, whereas the Neural Network offered improved capacity in capturing high-order dependencies. Additionally, eliminating feature transformations caused a substantial reduction in predictive accuracy, highlighting their role in mitigating feature skewness and amplifying meaningful patterns.

The test time prediction exhibited a similar trend. The full ensemble attained an accuracy of 96.80\%, as shown in Fig.~\ref{fig:ablationmetric}(b). The absence of any single base model produced noticeable deterioration in results, while the removal of feature engineering led to the most pronounced loss of performance. This observation reflects the intricate dependency between test time behavior and variables such as memory configuration, clock frequency, and test cycles. The Ridge Regression meta-learner, incorporated for its L2 regularization capability, played a significant role in improving generalization and stabilizing predictions across varying input distributions. In contrast, the Gradient Boosting Regressor meta-learner in the area prediction model contributed by refining residual errors produced by the base learners.

These results affirm that both the diversity of the ensemble and the application of tailored feature engineering are necessary to ensure model reliability and scalability. The findings suggest that simplified or partial variants of the proposed framework may be insufficient for capturing the complexities inherent in MBIST area and test time estimation, particularly when applied to large-scale or heterogeneous memory IPs.

\subsection{Comparative Analysis of Model Efficiency and Deployment Tradeoffs}
\label{sec:efficiency_tradeoff}
This section presents a comparative analysis of the performance efficiency and deployment trade-offs for four models used in predicting MBIST area and test time, namely XGBoost, LightGBM, Neural Network, and stacked ensemble. The models are evaluated based on key performance metrics, which include training time, inference time, model complexity, accuracy, and model size. These metrics are crucial for understanding both the predictive performance and the computational demands associated with deploying each model in practical applications.

XGBoost and LightGBM demonstrate relatively fast training and inference times, with training durations significantly lower than those of the Neural Network and stacked ensemble models. Their simpler architectures contribute to reduced complexity and memory usage, making them well-suited for resource-constrained environments. On the other hand, the Neural Network, due to its deep architecture \cite{b35}, requires more substantial training time and exhibits higher model complexity. Despite its higher resource consumption, the Neural Network model offers lower accuracy compared to the ensemble models, with a larger memory footprint that may limit its practical deployment in scenarios with limited resources.

\begin{figure}[!t]
  \centering
  \subfloat[]{
    \includegraphics[width=0.47\linewidth]{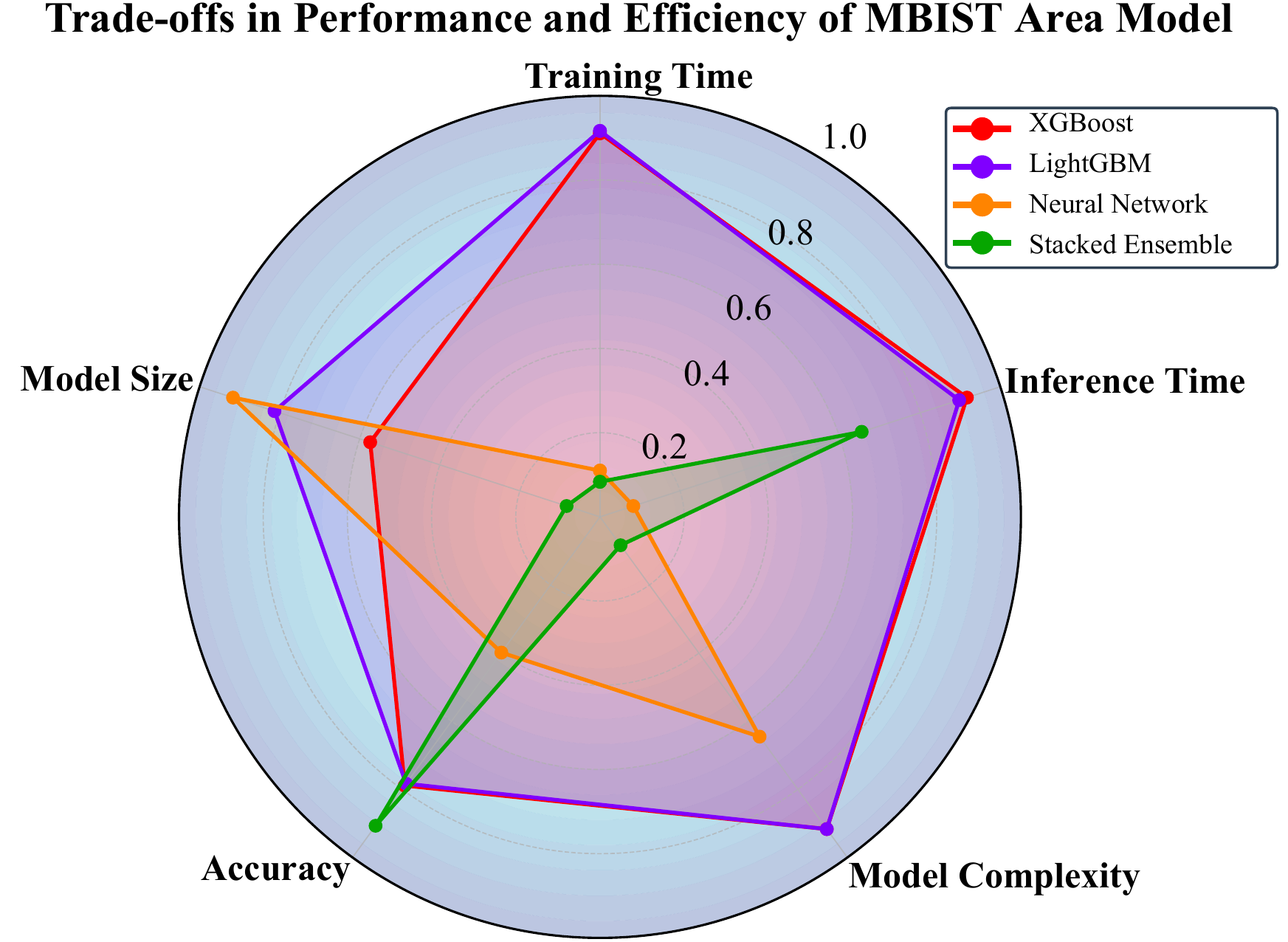}
  }
  \hfill
  \subfloat[]{
    \includegraphics[width=0.47\linewidth]{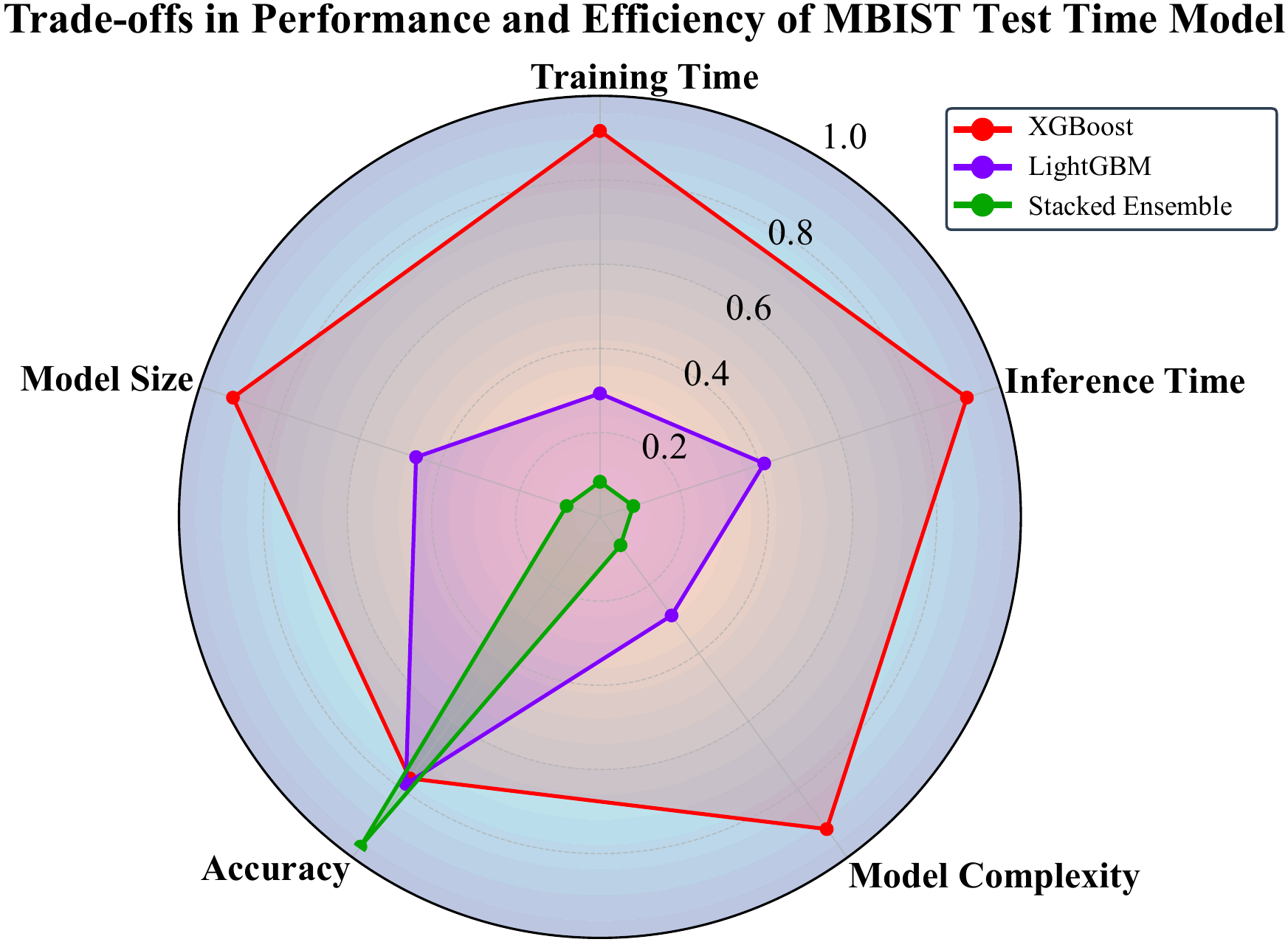}
  }
  \caption{Comparison of the trade-offs in performance and efficiency for the MBIST (a) area and (b) test time model across different models based on training time, inference time, model complexity, accuracy, and model size, where higher values indicate better performance.}
  \label{fig:modeltradeoff}
\end{figure}

The stacked ensemble, combining XGBoost, LightGBM, and Neural Network, provides the highest accuracy, achieving 90.68\% for MBIST area prediction. However, this performance comes at the cost of increased training time, inference time, and model complexity. While the stacked ensemble’s model size is larger due to the aggregation of base models and the meta-learner, its improved accuracy makes it an appealing choice for applications where prediction precision is prioritized. Nevertheless, the increased computational overhead may be acceptable, given that tasks like MBIST area and test time prediction do not require the same stringent timing constraints as real-time tasks such as video or image processing. Fig.~\ref{fig:modeltradeoff} illustrates the performance efficiency trade-offs between the models, highlighting the speed advantages of XGBoost and LightGBM, while also demonstrating the superior accuracy of the stacked ensemble and Neural Network models, with the stacked ensemble offering the best balance of accuracy and computational cost.

The performance metrics for each model in terms of training time, inference time, model complexity, accuracy, and model size for both MBIST area and test time prediction are summarized in Tables~\ref{tab:area_performance} and~\ref{tab:testtime_performance}. These tables show that while XGBoost and LightGBM exhibit relatively fast training and inference times, the stacked ensemble model, despite its larger size and increased computational cost, achieves significantly better accuracy. Specifically, the stacked ensemble achieves 90.68\% accuracy in predicting MBIST area and 96.80\% for test time, highlighting its superior performance in comparison to the other models.

\begin{table}[!t]
\centering
\caption{Comparison of Model Performance Metrics for MBIST Area Prediction}
\label{tab:area_performance}
\begin{tabular}{l cccc}
\toprule
\textbf{Metric} & \textbf{XGBoost} & \textbf{LGBM} & \textbf{NN} & \textbf{Stacked Ensemble} \\
\midrule
Train Time (s)     & 2.3752  & 0.8217  & 105.8558 & 109.6995 \\
Inference Time (s) & 0.0028  & 0.0083  & 0.2981   & 0.1293   \\
Complexity         & 108,800 & 108,800 & 214,017  & 431,817  \\
Accuracy (\%)      & 78.90   & 78.34   & 39.84    & 90.68    \\
Model Size (MB)    & 2.496   & 1.324   & 0.816    & 4.900    \\
\bottomrule
\end{tabular}
\end{table}

\begin{table}[!t]
\centering
\caption{Comparison of Model Performance Metrics for MBIST Test Time Prediction}
\label{tab:testtime_performance}
\begin{tabular}{l ccc}
\toprule
\textbf{Metric} & \textbf{XGBoost} & \textbf{LGBM} & \textbf{Stacked Ensemble} \\
\midrule
Train Time (s)     & 0.1753 & 0.8729 & 1.0737 \\
Inference Time (s) & 0.0014 & 0.0144 & 0.0128 \\
Complexity         & 11,504 & 46,500 & 58,006 \\
Accuracy (\%)      & 69.11  & 64.80  & 96.80  \\
Model Size (MB)    & 1.021  & 2.263  & 3.284  \\
\bottomrule
\end{tabular}
\end{table}

\section{Conclusion}
This study presents a stacked ensemble framework integrating XGBoost, LightGBM, and Neural Networks for the early prediction of MBIST area and test time in complex memory IPs. The approach employs advanced feature engineering techniques, including polynomial expansion and log transformation, to capture complex non-linear relationships and reduce data skewness, thereby improving model generalization. The ensemble model, using Gradient Boosting Regressor for area prediction and Ridge Regression for test time prediction, achieves high accuracies of 90.68\% and 96.80\%, respectively, outperforming traditional approaches such as linear regression and single-model Neural Networks. Semantic port recognition is automated to enhance feature extraction and scalability across diverse memory configurations. This method reduces the need for full synthesis and test pattern generation, thus accelerating the design process and lowering development costs. Future work will focus on integrating the proposed framework into existing EDA tools, along with expanding to additional performance metrics and traditional MBIST optimization techniques, to further enhance its practicality and robustness in real-world design flows.

\section*{Acknowledgments}
This work was supported by Professional Development Research University Special Grant number R.J130000.7113.07E51 MOHE and in part by MoHE Research Industry-Infused Incubator (MRI3). The authors would like to thank Intel Microelectronics (M) Sdn. Bhd. and MRI3 for their collaboration and technical support throughout this work.

\section*{Ethics and Privacy Statement}
This study does not involve human participants, animals, or personal data. The datasets used in this work consist solely of engineering design parameters and tool-generated measurements, and therefore do not raise ethical or privacy concerns.

\end{document}